\documentclass[aps,prl,twocolumn,longbibliography,amsmath,amssymb,nofootinbib,longbibliography,10pt]{revtex4-2}
\usepackage{amsfonts,amssymb,amsmath}
\usepackage{graphics}
\usepackage{graphicx}
\usepackage{nicefrac}
\usepackage{cases}
\usepackage{mathrsfs}
\usepackage{enumerate}
\usepackage{mathtools}
\usepackage{textcomp}
\usepackage{verbatim}
\usepackage{bm}          
\usepackage{soul}
\usepackage{cancel}
\usepackage{upgreek}
\usepackage{txfonts}
\usepackage{color}
\usepackage[colorlinks=true, urlcolor=blue, linkcolor=blue]{hyperref}
\usepackage[papersize={8.5in,11in}]{geometry}
\usepackage{esint}
\newcommand{\be}{\begin{equation}}
\newcommand{\ee}{\end{equation}}
\usepackage{makecell}

\begin{document}

\title{Wigner crystallization at large fine structure constant}
\author{Sandeep Joy}
\author{Brian Skinner}
\affiliation{Department of Physics, Ohio State University, Columbus, OH 43210, USA}
\date{\today}
\begin{abstract}
We consider the fate of the Wigner crystal state in a two dimensional system of massive Dirac electrons as the effective fine structure constant $\alpha$ is increased. In a Dirac system, larger $\alpha$ naively corresponds to stronger electron-electron interactions, but it also implies a stronger interband dielectric response that effectively renormalizes the electron charge. We calculate the critical density and critical temperature associated with quantum and thermal melting of the Wigner crystal state using two independent approaches. We show that at $\alpha \gg 1$, the Wigner crystal state is best understood in terms of logarithmically-interacting electrons, and that both the critical density and the melting temperature approach a universal, $\alpha$-independent value. We discuss our results in the context of recent experiments in twisted bilayer graphene near the magic angle.
\end{abstract}
\maketitle

\begin{figure}[htb]
\centering
\includegraphics[width=1.0 \columnwidth]{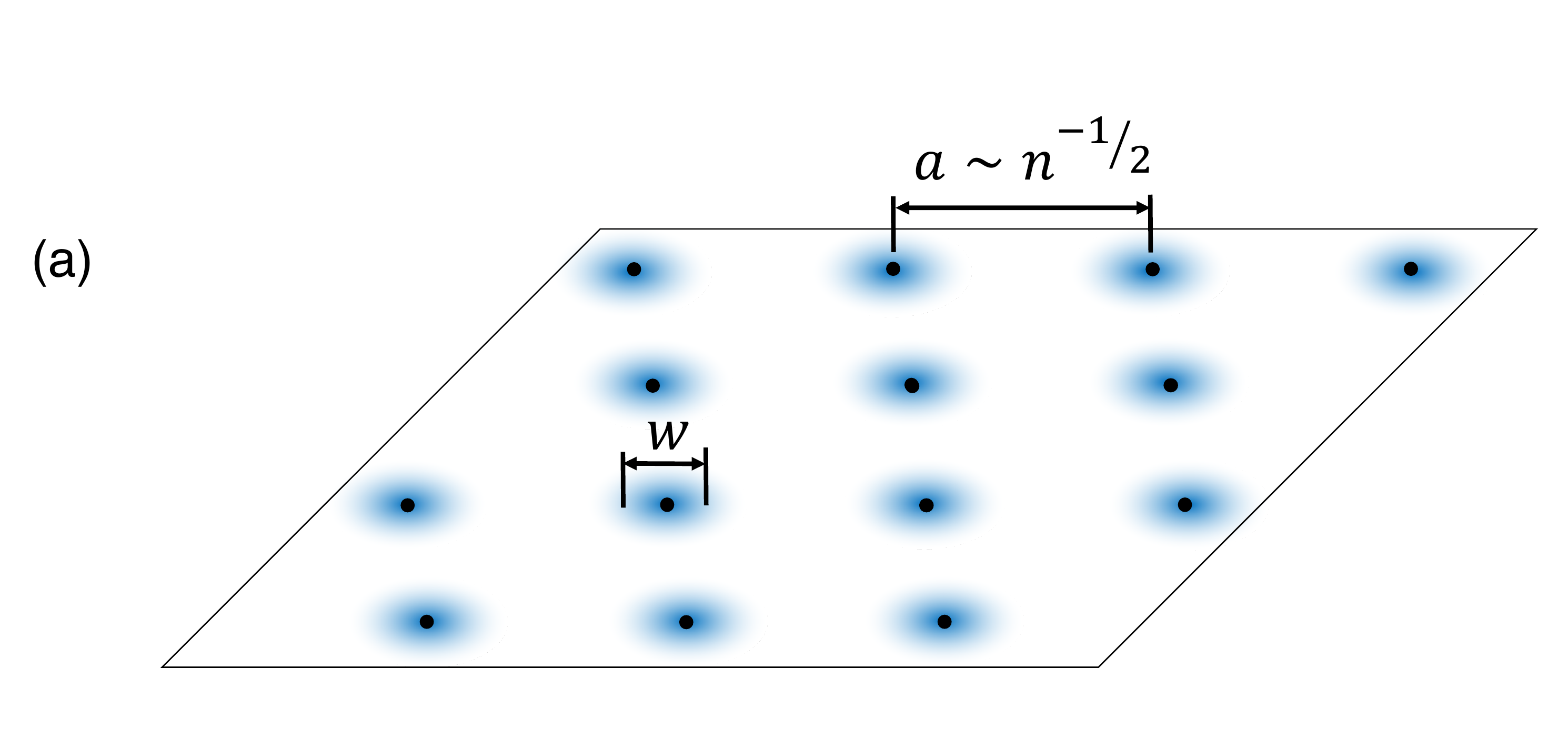}
\includegraphics[width=1.0 \columnwidth]{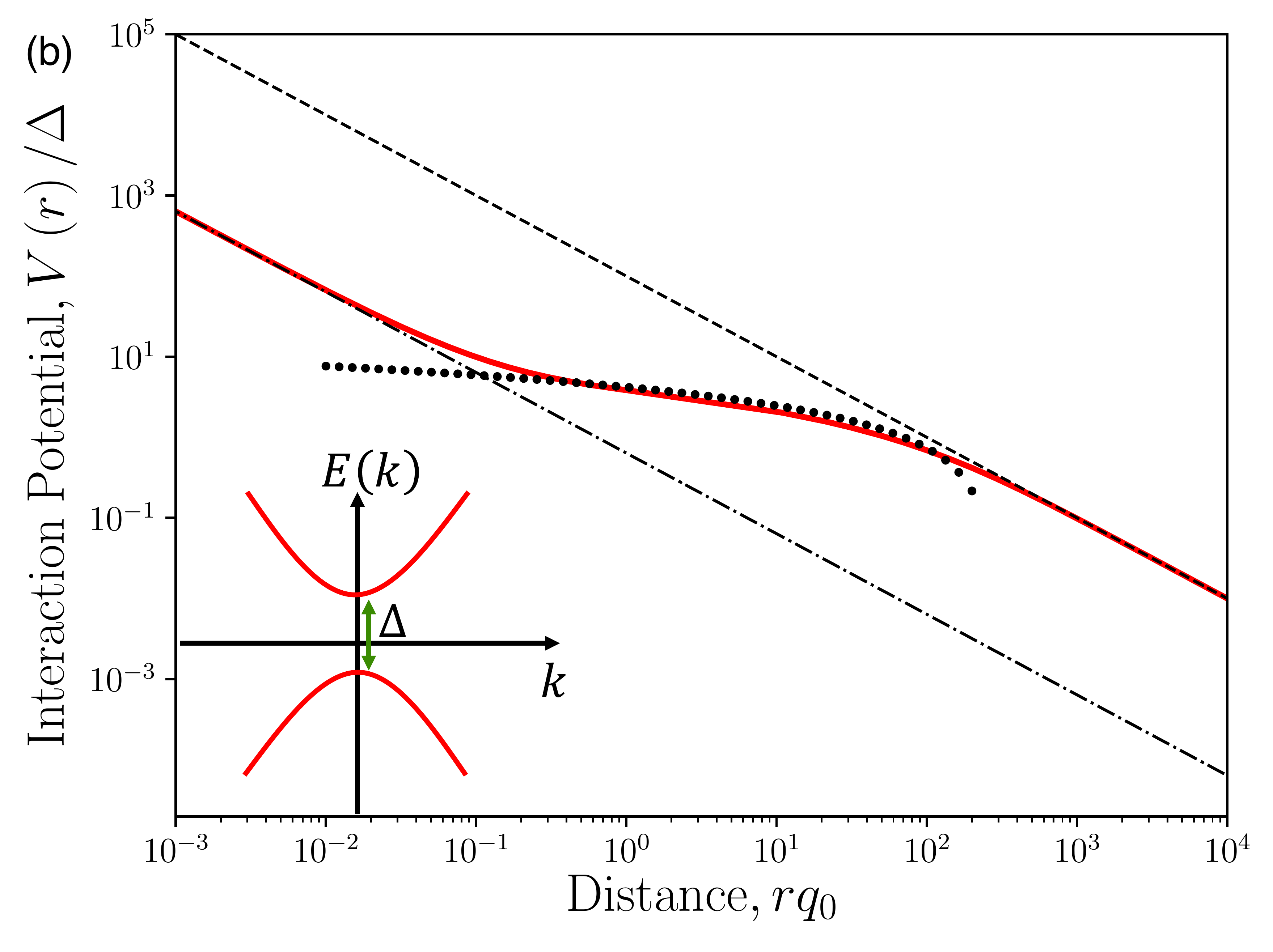}\label{fig:VScreen}
\caption{(a) Schematic picture of a WC, showing the lattice constant $a$ of the Wigner lattice and the typical width $w$ of the electron wavepacket. 
(b) The interaction potential $V(r)$ in real space, calculated (as an example) for $\alpha=100$. The three thin lines (dashed/dotted/dashed-dotted) correspond to the three limiting cases in Eq.~\ref{eq: vscreen}. The inset shows, schematically, the gapped Dirac dispersion relation $E(k)$.}
\label{fig:WC_Lattice}
\end{figure}

When the Coulomb interaction between electrons in a metal is sufficiently large compared to the electrons' kinetic energy, the electron system spontaneously breaks the continuous translational symmetry and forms a crystalline arrangement of electrons [illustrated in Fig.~\ref{fig:WC_Lattice}(a)] called a Wigner crystal (WC) \cite{Wigner_On_1934}. In a two dimensional (2D) electron gas at low temperature, this crystallization typically occurs in the limit of low electron density $n$, since the interaction between neighboring electrons scales as $n^{1/2}$ while the Fermi energy scales as $n$ for a dispersion relation with finite mass. 
Consider, for example, the illustrative case of electrons with a gapped Dirac Hamiltonian $H_0 = \hbar v\vec{\sigma}\cdot\vec{k}+(\Delta/2)\sigma_{z}$, which has a corresponding dispersion relation
\be 
E\left(k\right)=\sqrt{\left(\hbar vk\right)^{2}+\left(\Delta\big/2\right)^{2}}.
\label{eq:dirac}
\ee 
Here, $v$ is the Dirac velocity, $\Delta$ is the band gap, $\vec{k}$ is the wave vector, and $\vec{\sigma}$ represents the vector of Pauli matrices. When a small concentration of electrons is added to this system, the Fermi energy (relative to the band edge) is $E_F \sim \hbar^2 v^2 n / \Delta$, while the interaction energy is of the order $E_C \sim e^2 n^{1/2}/\epsilon_r$. (Here, $\hbar$ is the reduced Planck constant and $\epsilon_r$ is the dielectric constant. We use Gaussian units throughout this paper). Thus, the limit $E_C/E_F \gg 1$ that produces Wigner crystallization corresponds to  $n \ll \alpha^{2}\left(\Delta\big/\hbar v\right)^{2}$, where $\alpha=e^{2}\big/\epsilon_r \hbar v$ is the effective fine structure constant.

The fine structure constant $\alpha$ is the dimensionless parameter that typically characterizes the electron-electron interaction strength in Dirac systems. It is therefore no surprise that, as indicated above, increasing $\alpha$ (say, by decreasing the Dirac velocity $v$ or the dielectric constant $\epsilon_r$) leads to an increase in the prominence of the WC state. However, it is also known that increasing $\alpha$ leads to an increasing renormalization of the dielectric constant, due to screening by interband excitations. In a gapless 2D Dirac system like graphene, for example, this renormalization is such that $\epsilon_r \rightarrow \epsilon_r (1 + \pi \alpha/2)$ \cite{Ando_Screening_2006, Gorbar2002Magnetic}. The renormalization of the dielectric constant suggests an interesting question about the fate of the Wigner crystal state in the large-$\alpha$ limit. Namely: does the Wigner crystal state continue to become more prominent as $\alpha$ is increased far past $\alpha \sim 1$, or does the strong renormalization of the Coulomb interaction produce a qualitative change to the simple argument above? What is the nature of the Wigner crystal state at $\alpha \gg 1$?

Historically, questions about large fine structure constant have been largely hypothetical, since quantum electrodynamics in vacuum is famously characterized by a small fine structure constant $\alpha_\textrm{QED} \approx 1/137$. Monolayer graphene represents a platform for which the fine structure constant is of order unity or smaller, $\alpha_\textrm{graphene} \approx 2.2/\epsilon_r$,\footnote{Monolayer graphene does not permit WC phases since it has $\Delta = 0$ \cite{Dahal2006Absence}. On the other hand, Bernal-stacked bilayer graphene can exhibit WC phases when there is a perpendicular electric field \cite{Silvestrov2017Wigner}, which opens a band gap.} and theories based on small $\alpha$ are typically quite accurate quantitatively \cite{hofmann_why_2014, kolomeisky_optimal_2015}.
On the other hand, twisted bilayer graphene (TBG) offers a prominent example of a 2D Dirac system for which the effective fine structure constant may be very large, due to a nominal vanishing of $v$ near certain ``magic angles'' of relative twist (see, e.g., Refs.~\cite{Bistritzer_moire_2011, CastroNeto_topologically_2011, dosSantos_continuum_2012, zou_band_2018, fu_maximally_2018, kaxiras_exact_2019}). Motivated by this situation, in this paper we consider the generic question of Wigner crystallization at large $\alpha$. We comment specifically on TBG at the end of the paper.

Generally speaking, the screening of the Coulomb interaction between electrons at low frequency is described by the static dielectric function $\epsilon\left(q\right)$ (where $q$ is the wave vector), such that the (Fourier-transformed) screened potential is given by $\tilde{V}\left(q\right) = (2 \pi e^2/q)/\epsilon\left(q\right)$. The form of $\epsilon\left(q\right)$ resulting from interband transitions in a system with the gapped Dirac Hamiltonian $H_0$ has been calculated within the random phase approximation \cite{Kotov_Polarization_2008, Gorbar2002Magnetic}:
\be
\epsilon\left(q\right)=\epsilon_{r}\left(1+\alpha\left\{ \frac{q_{0}}{q}+\left(1-\frac{q_{0}^{2}}{q^{2}}\right)\arctan\left[\frac{q}{q_{0}}\right]\right\} \right).
\label{eq: epsq}
\ee
Here, the typical momentum scale $q_0 \equiv \Delta/\hbar v$ is equal to half the inverse Compton wavelength. The random phase approximation is justified by a $1/N$ expansion, where $N$ is the number of fermion flavors ($N = 8$ for twisted bilayer graphene), but it is not perturbative in $\alpha$ \cite{Kotov2012Electron}.

In real space, the above expression for $\epsilon\left(q\right)$ implies a potential $V(r)$ that has three asymptotic regimes at $\alpha \gg 1$:
\begin{equation}
V\left(r\right) \simeq \begin{cases}
2\hbar v/\pi r, & \qquad r\ll 1/q_0 \\
\frac{3}{4} \Delta  \ln \left[\frac{8\alpha}{3q_{0}r}\right], & \qquad 1/q_0 \ll r\ll \alpha / q_0 \\
e^2/\epsilon_r r, & \qquad r \gg \alpha / q_0 .\\
\end{cases}
\label{eq: vscreen}
\end{equation}
This expression for $V(r)$ is reminiscent of the interaction energy between two point charges embedded in a slab of large dielectric constant $\epsilon_r \sim \alpha$, for which confinement of electric field lines within the slab produces a logarithmic variation of the potential at intermediate distances \cite{keldysh1964infrared, rytova_the_1967, keldysh_coulomb_1979}. (A similar logarithmic interaction arises between vortices in type-II superconductors \cite{ketterson_Superconductivity_1999}.) 
Notice that at short distances $r \ll 1/q_0$ the interaction potential becomes independent of the electron charge. Effectively, at such short distances the value of $e^2$ is renormalized toward $\sim e^2/\alpha$ by the interband dielectric response, as in the problem of the ``supercritical'' impurity charge \cite{Pomeranchuk_Smorodinsky_1945, Zeldovich_Popov_1972, fogler_hypercritical_2007, shytov_vacuum_2007, shytov_atomic_2007, Pereira_Supercritical_2008, gamayun_supercritical_2009, yang_observing_2013}.
The potential $V(r)$ is plotted in Fig.~\ref{fig:WC_Lattice}(b), along with the asymptotic expressions of Eq.~(\ref{eq: vscreen}), for the case $\alpha = 100$.

It is important to note that, in general, one can expect a significant renormalization of the ``bare'' dispersion relation when $\alpha$ is large. For example, at small $\Delta$ and large $\alpha$ one would naively expect an excitonic instability, in which the electron-hole binding energy becomes larger than the band gap, leading to a condensation of excitons and the emergence of a larger, many-body gap.  Whether such an excitonic instability actually occurs is a subtle question, which involves careful consideration of the self-consistent screening (see, e.g., Refs.~\onlinecite{janssen2016excitonic, Janssen2017phase}).  Generally speaking, however, large $\alpha$ produces significant interaction-induced renormalization of the band structure, as discussed in detail elsewhere (for example, Refs.~\cite{Kotov2012Electron, song_electron_2013}). In this paper we take as given that a dispersion relation exists for low-energy quasiparticles that can be described by Eq.~(\ref{eq:dirac}), even when $\alpha$ is large. But the parameters $\Delta$ and $v$ in this equation should be viewed as the interaction-renormalized values, rather than the values that would correspond to a hypothetical non-interacting system. The effective value of $\alpha$ corresponding to a given non-interacting band structure is a nontrivial question, since the band velocity is generally renormalized upward by strong interactions. In TBG, empirical measurements of the Dirac velocity range between $7 \times 10^{4}$\,m/s and $2 \times 10^5$\,m/s near the magic angle \cite{Polshyn_large_2019, cyprian_correlation_2021, cyprian2021spectroscopic}. Such velocities are significantly larger than the predictions of noninteracting theories, but still produce an estimate for $\alpha$ in the range $2$ -- $6$ (using $\epsilon_r \approx 5$, which corresponds to a hexagonal boron nitride substrate). 

The standard semiclassical picture of the WC phase is that electrons are arranged in a triangular Wigner lattice, with each electron residing in the potential well created by repulsion with its neighbors. In this sense one can imagine that, deep within the WC phase, each electron effectively constitutes a harmonic oscillator (HO) in a locally parabolic potential. At very low electron density $n$, the radius $w$ of the HO ground state wavefunction is much smaller than the lattice constant $a$. However, as the density is increased, the ratio $\eta = w/a$ increases. The Lindemann criterion of melting states that at a critical value of $\eta = \eta_c \approx 0.23$ \cite{Babadi_Universal_2013, Astrakharchik_Quantum_2007}, the electron system undergoes a phase transition from a WC to a Fermi liquid (FL) state.

Below we estimate the critical density associated with this melting transition using two complementary calculations. First, we use the HO description (which is asymptotically exact at small $n$) to analytically calculate the Lindemann ratio deep within the WC phase. This approach allows us to estimate both the critical density $n_c$ and the critical temperature $T_m$ associated with melting. Second, we perform a Hartree calculation of the total energy using the many-body Hamiltonian and a variational wavefunction, which gives us a separate estimate of $n_c$.

In order to produce an analytical estimation of the Lindemann ratio deep within the WC phase, we calculate the parabolic coefficient of the confining potential $U(r)$ for an electron centered at the origin $(r=0)$. We use the standard description for electrons deep in the WC state, in which all other electrons are treated as point charges residing at points on the Wigner lattice \cite{mahan1990many}. The potential $U(r)$ can be Taylor expanded as
\be
U\left(\vec{r}\right)-U(0) \simeq \frac{r^{2}}{4}\sum_{i\neq0}\left[V''\left(\left|\vec{R}_{i}\right|\right)+\frac{V'\left(\left|\vec{R}_{i}\right|\right)}{\left|\vec{R}_{i}\right|}\right],
\label{eq: ho}
\ee
where $\vec{R}_i$ denotes the position of the Wigner lattice point with index $i$.

The harmonic oscillator frequency $\omega$ is defined by ${U(r) - U(0) = \frac{1}{2}m \omega^2 r^2}$, where $m = \Delta/2v^2$ is the electron mass at low energy. 
In the limit $\alpha \ll 1$ (weak coupling limit), the interaction potential is essentially unmodified by the interband dielectric response, and $V(r) \simeq e^2/\epsilon_r r = \alpha \hbar v/r$.  In this case, Eq.~(\ref{eq: ho}) gives
\be
m\omega^{2}=\frac{\alpha\hbar v\lambda}{2a^{3}},
\ee
where $\lambda \approx 11.03$ is a numerical constant and $a = (\sqrt{3}n / 2)^{-1/2}$ is the lattice constant of the Wigner lattice. Using the width of the corresponding ground state of a 2D HO, $\sqrt{\langle r^2 \rangle} =\left(\hbar\Big/m\omega\right)^{1/2}$, the Lindemann ratio can be computed to be 
\be
\eta=\left(\frac{8\sqrt{3}}{\lambda^{2}}\right)^{1/8}\left(\frac{n}{\alpha^{2}q_{0}^{2}}\right)^{1/8}, \hspace{5mm} (\alpha \ll 1).
\label{eq: lind_small}
\ee
This result is similar to the original stability calculation using the Wigner-Seitz approximation \cite{mahan1990many}. 

In the opposite limit of $\alpha\gg1$ (strong coupling limit) and at densities  $1/\alpha^{2}\ll n/q_{0}^{2}\ll1$, the interaction relevant to WC formation becomes logarithmic in nature (see Eq.~\ref{eq: vscreen}). Inserting such a logarithmic interaction directly into Eq.~\ref{eq: ho} does not give a finite HO frequency, due to Earnshaw's theorem \cite{purcell2013electricity}. Instead, understanding the stability of the WC phase in this limit requires one to think explicitly about the background charge, as pointed out in the original paper by Wigner for three dimensions \cite{Wigner_On_1934}. We account for the background charge by replacing the logarithmic interaction with the screened Coulomb interaction in 2D, given by
\be
V\left(r\right)=\lim_{\kappa\rightarrow0}\frac{3\Delta}{4}K_{0}\left[\frac{\kappa r}{a}\right].
\ee
Here $K_{0}$ is the modified Bessel function of the second kind and $\kappa$ is the dimensionless screening parameter. In this description each electron is surrounded by a ``screening cloud'' of compensating positive charge, which has a characteristic radius $a/\kappa$. The physical case of uniform background charge corresponds to the limit where neighboring screening clouds are strongly overlapping, $\kappa \ll 1$.
Inserting this expression for $V(r)$ into Eq.~(\ref{eq: ho}) and taking the limit $\kappa \rightarrow 0$ after performing the sum gives
\be
m\omega^{2}=\frac{3 \pi}{4} n  \Delta.
\ee
The corresponding Lindemann ratio is
\be
\eta=\left(\frac{2}{\pi}\right)^{1/4}\left(\frac{n}{q_{0}^{2}}\right)^{1/4}, \hspace{5mm} (\alpha \gg 1).
\label{eq: lind_large}
\ee
Thus, in the strong coupling limit, the Lindemann ratio $\eta$ is independent of $\alpha$. This independence can be seen as a consequence of the renormalization of $e^2/\epsilon_r$ toward $\hbar v$, as mentioned above. 

\begin{figure}[htb]
\centering
\includegraphics[width=1.0 \columnwidth]{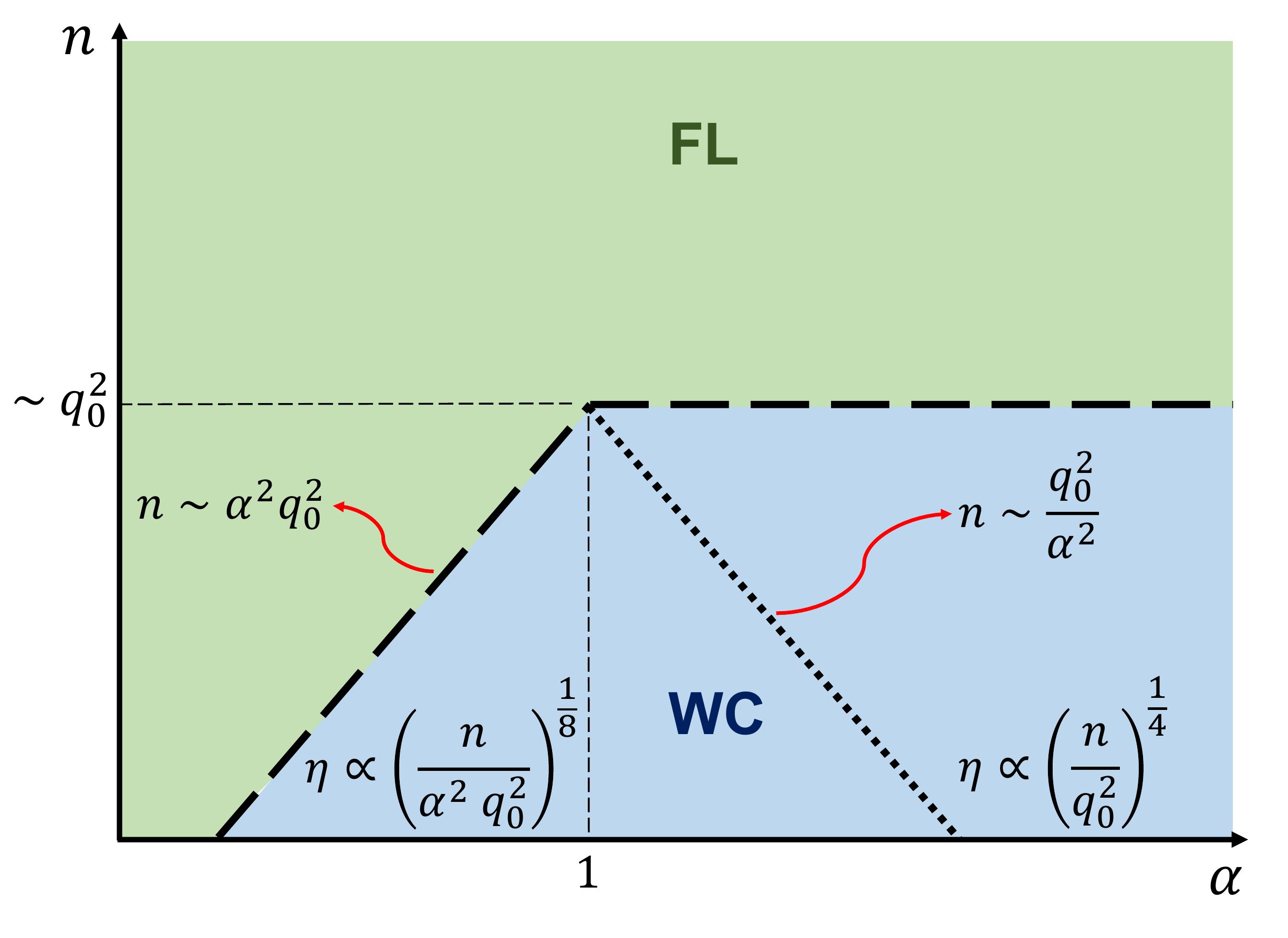}
\caption{Schematic phase diagram based on the Eqs.~\ref{eq: lind_small} and \ref{eq: lind_large} for the Lindemann ratio. ``FL'' denotes the Fermi liquid phase. Different WC regimes are labeled by the corresponding behavior of the Lindemann ratio $\eta$. Both axes are plotted in logarithmic scale.}
\label{fig: PD_Theo}
\end{figure}

We can produce an estimate for the quantum melting transition of the WC by extrapolating our results for $\eta$ to the critical value $\eta = \eta_c$. This procedure gives for the critical concentration
\begin{align}
n_c & \simeq C_1 \alpha^2 q_0^2, &  (\alpha \ll 1)
\label{eq: ncsmall} \\
n_c & \simeq C_2 q_0^2,  &  (\alpha \gg 1)
\label{eq: nclarge}
\end{align}
where $C_1$ and $C_2$ are numerical constants; naively setting $\eta = \eta_c$ in Eqs.~(\ref{eq: lind_small}) and (\ref{eq: lind_large}) gives $C_1 \approx 6\times10^{-5}$ and $C_2 \approx 4\times10^{-3}$. The corresponding phase diagram is shown schematically in Fig.~\ref{fig: PD_Theo}, along with asymptotic expressions for the Lindemann ratio.

\begin{figure}[htb]
\centering
\includegraphics[width=1.0 \columnwidth]{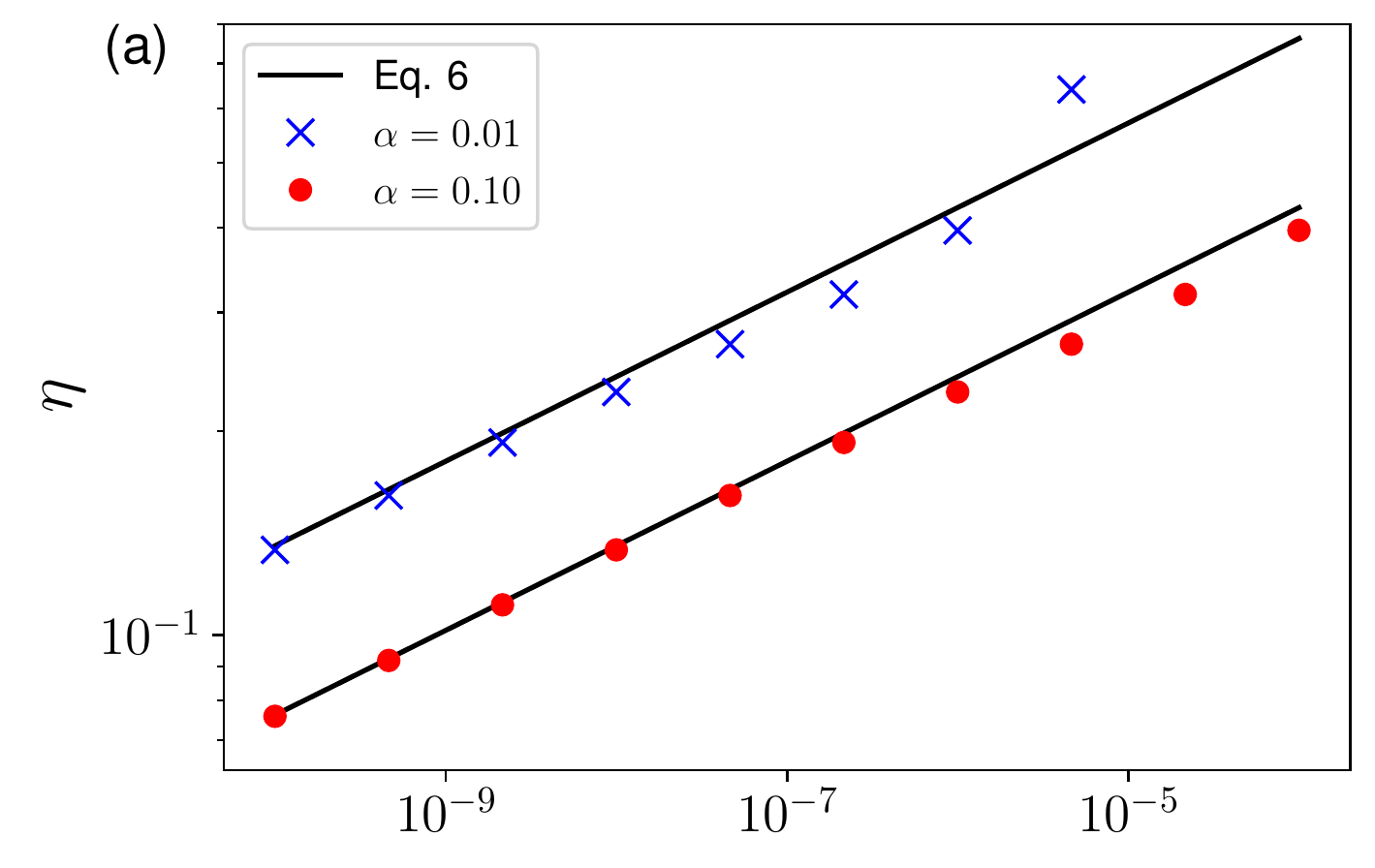}
\includegraphics[width=1.0 \columnwidth]{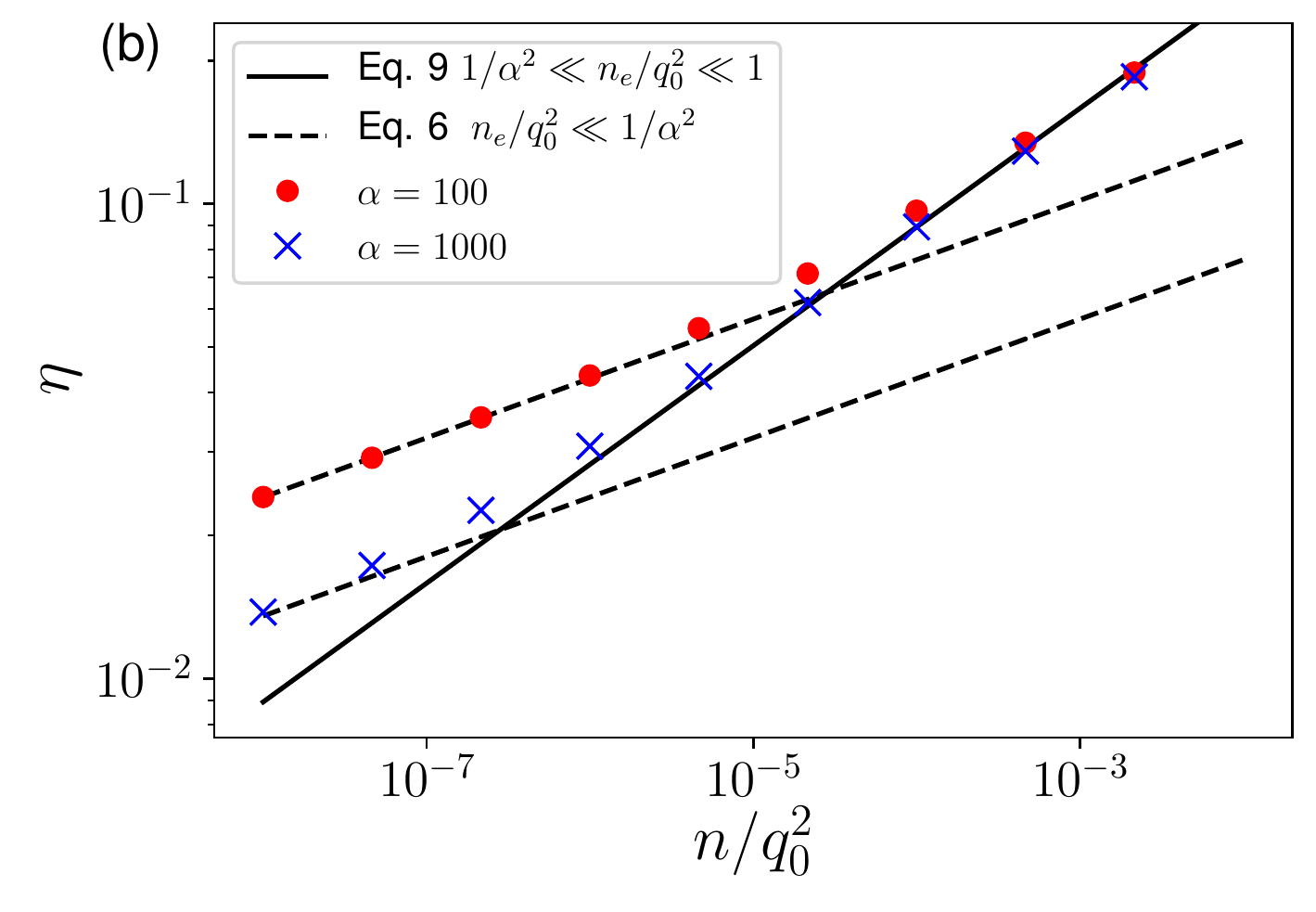}
\caption{Lindemann ratio as a function of electron density in the limit of (a) $\alpha \ll 1$ and (b) $\alpha \gg 1$. The symbols show numeric calculations using the variational approach, and lines show our analytical calculations using the HO approximation.}
\label{fig:Lind}
\end{figure}

An alternative method for calculating the Lindemann ratio of the WC state, which does not require the HO approximation, is to use the variational principle with a many-electron trial wavefunction. Specifically, we search for the state that minimizes the expectation value of the Hamiltonian $H = \sum_i \hat{E}_i + \frac{1}{2} \sum_{ij} V(r_{ij})$, where $\hat{E}_i$ is the kinetic energy operator for electron $i$ and $V(r)$ is the screened interaction.

Following Refs.~\cite{Skinner_Chemical_2016, skinner_interlayer_2016}, we choose a trial wavefunction that consists of Gaussian wavepackets centered around the points of the Wigner lattice. The width $w$ of each wavepacket is treated as a variational parameter. Since the exchange interaction between electrons is exponentially small deep in the WC regime, in this limit we can accurately approximate the electrostatic energy per electron using the Hartree approximation. The corresponding expressions for the kinetic and potential energy are derived in Supplemental Material \cite{SM}. The value of $w$ that minimizes the total energy gives an estimate of the Lindemann ratio, $\eta = w/a$. We perform this minimization numerically in order to calculate the Lindemann ratio as a function of $n$ and $\alpha$.

\begin{figure}[htb]
\centering
\includegraphics[width=1.0 \columnwidth]{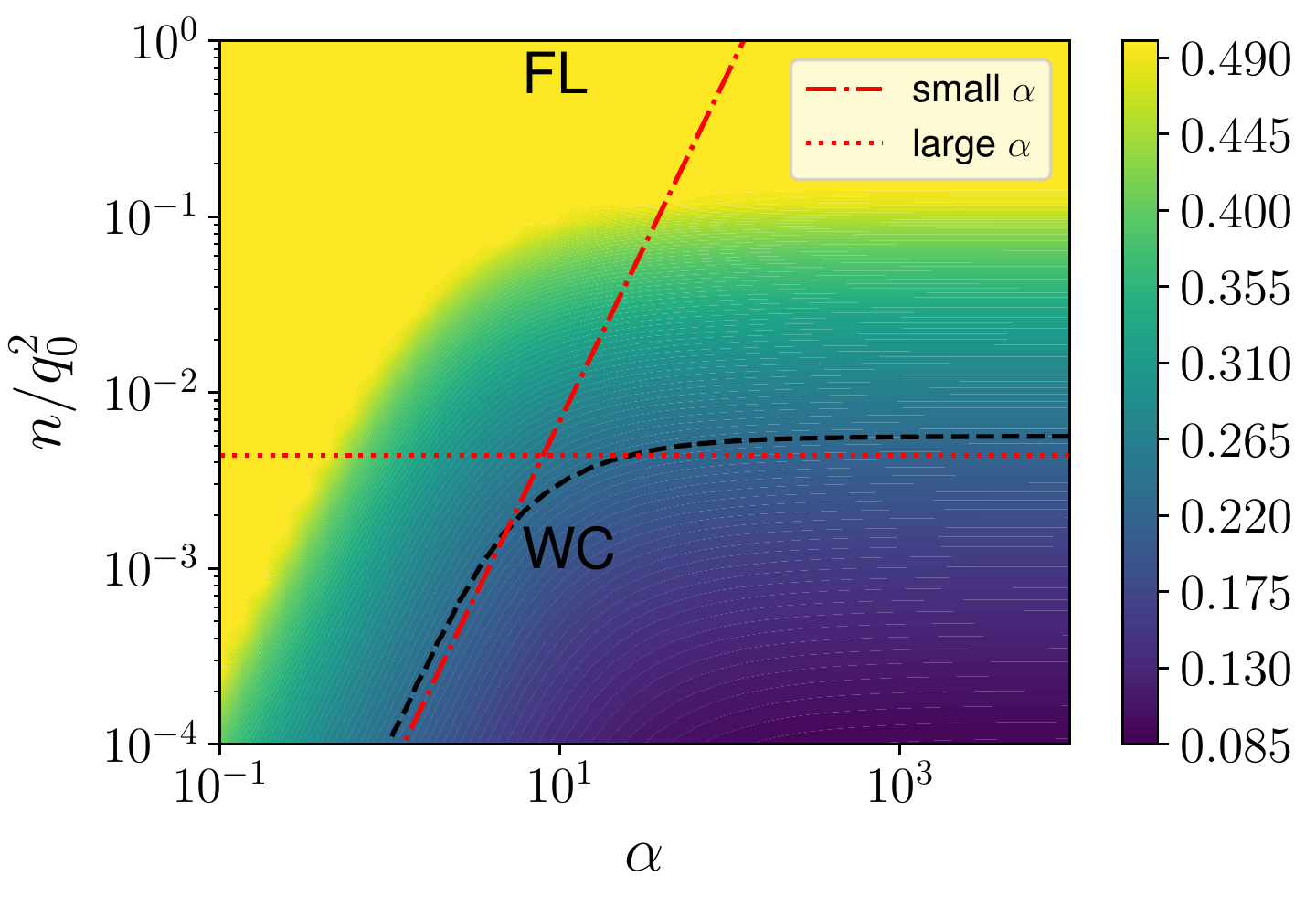}
\caption{Lindemann ratio as a function of both $\alpha$ and electron density, calculated numerically using the variational approach. The dashed black line corresponds to $\eta = \eta_c = 0.23$, and therefore represents an estimation of the phase boundary between FL and WC phases. The red dashed-dotted/dotted lines represents the analytical estimation for the critical density [Eqs.~(\ref{eq: ncsmall}) and (\ref{eq: nclarge})]. }
\label{fig: PD}
\end{figure}

The result of this numerical calculation is plotted in Fig.~\ref{fig:Lind} for a few example values of $\alpha \ll 1$ [Fig.~\ref{fig:Lind}(a)] and $\alpha \gg 1$ [Fig.~\ref{fig:Lind}(b)]. The numeric result for $\eta$ closely matches the analytical expressions presented in Eqs.~\ref{eq: lind_small} and \ref{eq: lind_large} in the appropriate limits. Our numeric result for $\eta$ is plotted as a function of both $\alpha$ and $n$ in Fig.~\ref{fig: PD}. An estimation of the phase boundary between the FL and WC phases is determined by setting the variational value of $\eta$ equal to $\eta_c$ (black dashed curve). This phase boundary is consistent with the analytical estimates (red lines, compare also with Fig.~\ref{fig: PD_Theo}). Of course, one should keep in mind that neither calculation is quantitatively accurate in the immediate vicinity of the phase transition, and so our result for the critical density $n_c$ should be viewed only as a scaling estimate. 

\begin{figure}[htb]
\centering
\includegraphics[width=1.0 \columnwidth]{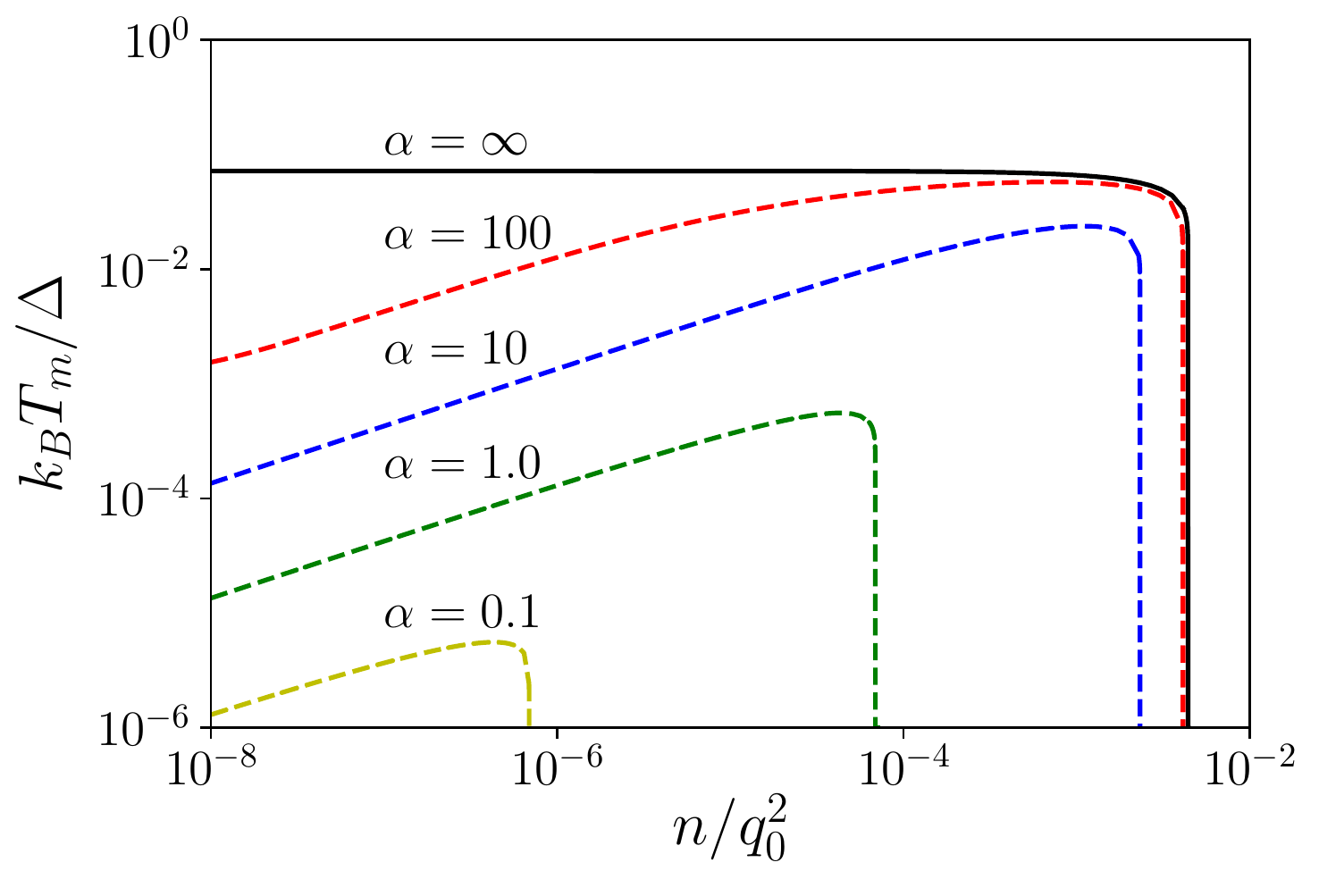}
\caption{The estimated melting temperature $T_m$ as a function of the electron density, plotted for a range of $\alpha$ values. In the limit $\alpha \rightarrow \infty$ (black curve), the melting temperature approaches a finite, universal value at $n \ll n_c$.}
\label{fig: Thermal_PD}
\end{figure}

So far we have confined ourselves to considering the Wigner crystal state at zero temperature. We now discuss the critical temperature $T_m$ associated with melting of the WC state. At finite temperature the HO description used above admits a simple generalization, in which the value of $\langle r^2 \rangle$ is replaced with the thermal expectation value. This expectation value can be calculated in a straightforward way as 
\be
\left\langle r^{2}\left(T\right)\right\rangle  =\left(\frac{\hbar}{m\omega}\right)\coth\left[\frac{\hbar\omega}{2 k_B T}\right],
\ee
where $k_B$ is the Boltzmann constant. Using our analytical result for the HO frequency $\omega$ (see Eq. \ref{eq: ho}), this expression enables us to estimate the melting temperature by setting $\eta_c=\sqrt{\langle r^{2}(T)\rangle}/a$. This procedure gives
\be 
k_B T_m \approx \min \left\{ A_{1} \frac{\alpha n^{1/2}\Delta}{q_{0}}, \ A_2 \Delta \right\},
\label{eq:Tm}
\ee 
at $n \ll n_c$, where $A_1$ and $A_2$ are numerical constants which we estimate as $A_1 \approx 0.136$, $A_2 \approx 0.072$. When the electron density $n$ approaches $n_c$, the melting temperature drops rapidly to zero.
Our numerical result for the melting temperature is presented in Fig.~\ref{fig: Thermal_PD}. Notably, the melting temperature approaches a constant, $n$-independent value at $\alpha \gg 1$ and $n < n_c$. 

In summary, in this paper we have considered both the quantum and thermal melting of the WC state in a gapped Dirac system at arbitrary values of the effective fine structure constant $\alpha$. We find two qualitatively different regimes, depending on whether $\alpha$ is large or small. The limit $\alpha \ll 1$ corresponds to the usual case of $1/r$-interacting electrons, and gives a critical density and a maximum melting temperature that both increase as $\alpha^2$. On the other hand, at $\alpha \gg 1$ the Wigner crystal state can be understood in terms of logarithmically-interacting electrons, with an overall interaction strength that is renormalized toward the band gap $\Delta$ by the interband dielectric response. This renormalization produces a universal, $\alpha$-independent critical density $n_c \propto q_0^2$, and a melting temperature $k_B T_m \propto \Delta$.

At present, the most prominent platform for testing these ideas is TBG \cite{cao_correlated_2018, Cao_unconventional_2018}, in which the moir\'{e} pattern created by the relative twist produces strong renormalization of the graphene band structure, leading to a small Dirac velocity and apparently large $\alpha$ (as discussed above).  Transport and scanning tunneling microscopy studies indicate small energy gaps at certain band fillings that are commensurate with the moir\'{e} pattern (see, e.g., Refs.~\onlinecite{Serlin_Intrinsic_2020, Stepanov_untying_2020, Park_Flavour_2021, Xie_spectroscopic_2019, choi2021interactiondriven}). For example, transport measurements (which produce a lower-bound estimate of the energy gap \cite{huang2021conductivity}) typically yield values of $\Delta$ between a few tenths of an meV and several meV \cite{Serlin_Intrinsic_2020, Stepanov_untying_2020, Park_Flavour_2021}, while scanning tunneling microscopy studies suggest a gap as much as ten times larger \cite{Xie_spectroscopic_2019, choi2021interactiondriven}. Following the logic of this paper, one can generically expect a WC state when the filling of electrons $n$ relative to the insulating value is very small. Since a WC is typically an insulator due to pinning by disorder, one should therefore expect that the insulating state occupies a finite range of electron density, even in the limit of very low disorder. 
Equation (\ref{eq: nclarge}) suggests that this range of density is of order $10^9$\,cm$^{-2}$ (using, as an estimate, $\Delta \approx 2$\,meV and $v \approx 10^5$\,m/s) in the limit of low temperature. At temperatures larger than the maximum value of $T_m$ (which we estimate as $\sim 1.6$\,K using Eq.~\ref{eq:Tm} and the same parameters as above), the WC state should melt for all densities, leading to a disappearance of the WC insulating state. Experimentally, the presence of the WC state can generally be inferred by the presence of a characteristic sharp ``pinning voltage'' in the $I$-$V$ curve \cite{Falson2021competition}. Tracking the onset of this pinning as a function of electron density and temperature yields a phase diagram which can be compared with Figs.~\ref{fig: PD_Theo} and \ref{fig: PD}.

Finally, we would like to draw a contrast between our description of the WC state and that of Refs.~\cite{Padhi_Doped_2018, Padhi_Pressure_2019}. We describe the WC state using an effective low-energy band structure, which in the context of TBG minibands requires that all relevant momenta are small compared to the inverse moir\'{e} lattice constant $\lambda_m$. Thus, our theory is applicable only to situations where the density of electrons, relative to the gap, is very small: $n \lambda_m^2 \ll 1$. Since our estimate for $n_c \ll 1/\lambda_m^2 \sim 10^{12}$\,cm$^{-2}$, the description is self-consistent for TBG. In contrast, Refs.~\cite{Padhi_Doped_2018, Padhi_Pressure_2019} use the language of Wigner crystallization to discuss situations where there are one or multiple electrons per moir\'{e} cell. We also point out that, while we have estimated the location of the WC-FL phase transition, this transition is generally more complicated than a simple first-order transition and involves a sequence of microemulsion phases across a narrow range of densities surrounding $n_c$ \cite{Spivak_Phases_2004}. Such phases are beyond the level of sophistication of our description. 
\acknowledgments
{\it Acknowledgments.-} We are thankful to Cyprian Lewandowski and J.~C.~W.~Song for useful discussions. This work was supported by the NSF under Grant No.~DMR-2045742.
\bibliography{WC}

\end{document}


\widetext
\begin{center}
\textbf{\large Supplementary Information for ``Wigner crystallization at large fine structure constant"}
\end{center}
\setcounter{equation}{0}
\setcounter{figure}{0}
\setcounter{table}{0}
\setcounter{page}{1}
\makeatletter
\renewcommand{\theequation}{S\arabic{equation}}
\renewcommand{\thefigure}{S\arabic{figure}}
\renewcommand{\bibnumfmt}[1]{[S#1]}
\renewcommand{\citenumfont}[1]{S#1}

\begin{center}
Sandeep Joy, Brian Skinner \\
\textit{Department of Physics, Ohio State University, Columbus, OH 43210, USA} \\
(Dated: \today)
\end{center}
\section{S1. Derivation of variational energy}
\label{sec: expec}

Here we derive the expression for the variational energy used in our estimation of the Lindemann ratio $\eta$. We calculate the Hartree energy of a variational wave function given the many-body Hamiltonian
\be
H = \sum_i \hat{E}_i + \frac{1}{2} \sum_{ij} V(r_{ij}),
\ee
where $\hat{E}_i$ is the kinetic energy operator for the $i^{th}$ electron and $V(r_{ij})$ is the screened Coulomb interaction between the $i^{th}$ and $j^{th}$ electrons. For a given variational wave function the expectation value of the Hamiltonian can be written as
\be 
\langle H \rangle_\textrm{var} = N ( K_\textrm{var} + U_\textrm{var} ),
\ee 
where $N$ is the total number of electrons.
Below we derive the expressions for the kinetic and potential energy per electron, $K_\textrm{var}$ and $U_\textrm{var}$, respectively.

\begin{figure}[htb]
\centering
\includegraphics[width=0.6 \columnwidth]{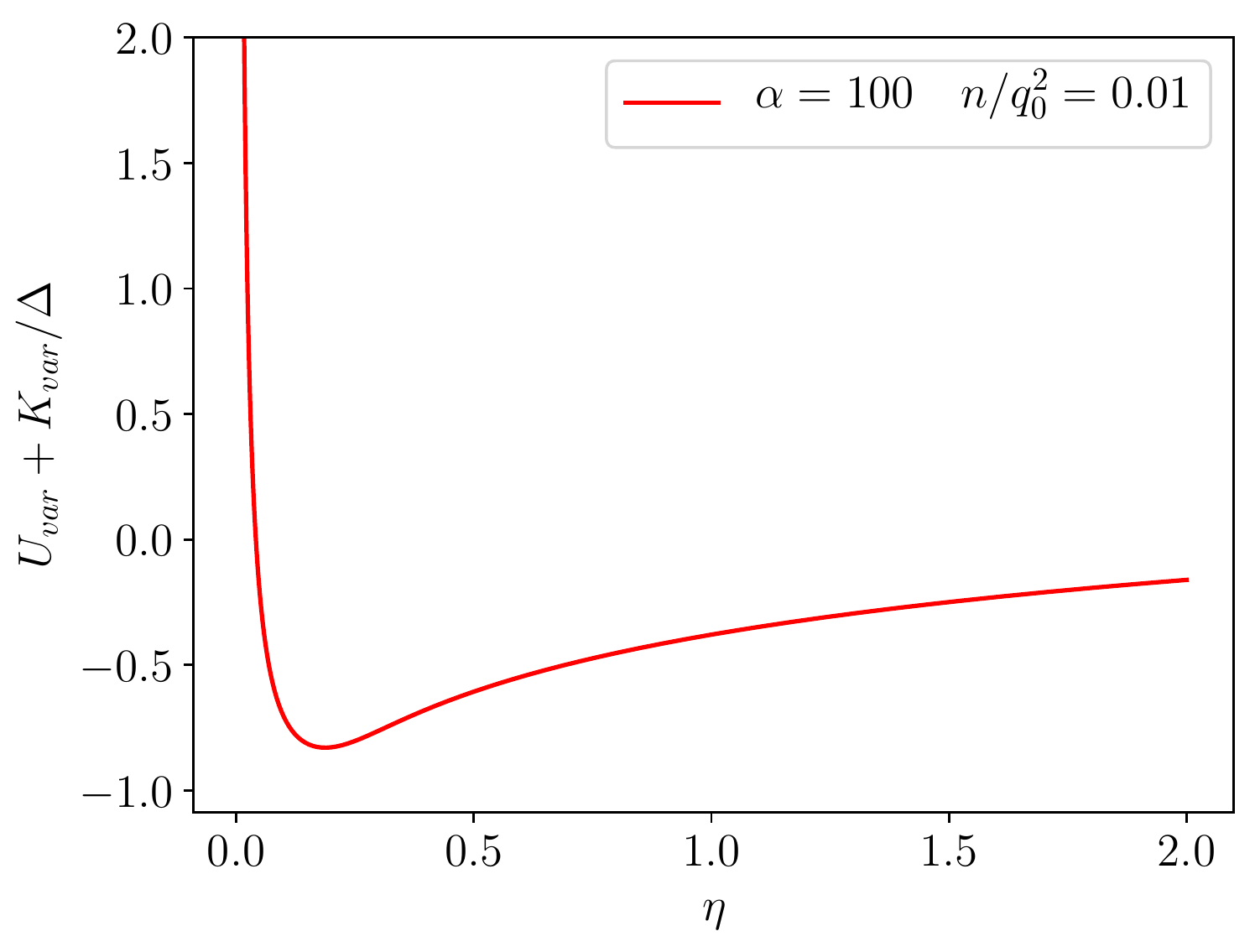}
\caption{Total variational energy, $U_{\text{var}}+K_{\text{var}}$, is plotted as a function of $\eta$ for the example case of $\alpha=100$ and $n/q_{0}^2=0.01$. The corresponding minimum of the total variational energy can be found numerically, and in this case gives $\eta \approx 0.187$.}
\label{fig: TE}
\end{figure}

We use a trial wavefunction consisting of Gaussian wavepackets centered around the points $\vec{R}_{i}$ of the Wigner lattice. A single wavepacket is described by
\be
\varphi_{i}\left(\vec{r}\right)=\frac{1}{\sqrt{\pi w^{2}}}\exp\left[-\frac{\left|\vec{r}-\vec{R}_{i}\right|^{2}}{2w^{2}}\right].
\label{eq: gaussian}
\ee 
The potential energy per electron can be evaluated to be
\begin{align}
    \begin{split}
        U_{\text{var}}&=\frac{1}{2}\int d\vec{r}\int d\vec{r'}V\left(r-r'\right)\left|\varphi_{0}\left(\mathbf{r}\right)\right|^{2}\sum_{i\neq0}\left|\varphi_{i}\left(\mathbf{r'}\right)\right|^{2},\\&=\frac{1}{2}\sum_{i\neq0}\int\frac{d\vec{q}}{\left(2\pi\right)^{2}}\tilde{V}\left(q\right)\rho_{0}\left(-\vec{q}\right)\rho_{i}\left(\vec{q}\right),\\&=\frac{n}{2}\sum_{\ell\neq0}\tilde{V}\left(\vec{K}_{\ell}\right)\exp\left[-\vec{K}_{\ell}^{2}w^{2}\right]-\frac{1}{2}\int\frac{d\vec{q}}{\left(2\pi\right)^{2}}\tilde{V}\left(q\right)\exp\left[-q^{2}w^{2}\right].
    \end{split}
\end{align}
Here the vectors $\vec{K}$ are the elements of reciprocal lattice space. In the first term on the final expression on right hand side, $\ell=0$ is removed in order to cancel the contribution from the uniform positive background. In terms of $n$, the magnitude of the reciprocal lattice vectors are given by
\be
\left|\vec{K}_{l,m}\right|=\left(\frac{4\pi}{\sqrt{2\sqrt{3}}}\right)n^{\frac{1}{2}}\sqrt{l^{2}+m^{2}-lm}.
\ee
The variational kinetic energy $K_\text{var}$ can be evaluated in momentum space as
\be
K_{\text{var}}=\int\frac{d\vec{q}}{\left(2\pi\right)^{2}}\left|\varphi_{0}\left(\vec{q}\right)\right|^{2}E\left(q\right).
\ee
For a gapped Dirac dispersion, where $E\left(k\right)=\sqrt{\hbar^{2}v^{2}k^{2}+\frac{\Delta^{2}}{4}}$, we can evaluate $K_{\text{var}}$ to be
\be
K_{\text{var}}=\Delta\left(\frac{1}{2}+\frac{\sqrt{2\pi}e^{\frac{q_{0}^{2}w^{2}}{2}}\text{erfc}\left(\frac{q_{0}w}{\sqrt{2}}\right)}{4q_{0}w}\right),
\ee
where $\textrm{erfc}$ denotes the complementary error function. The limiting behavior of the above expression can be calculated as
\be
K_{\text{var}} \approx\left\{ \begin{array}{cc}
\frac{1}{2q_0w}\sqrt{\frac{\pi}{2}}, & q_0w\longrightarrow0,\\
\frac{1}{2\left(q_0w\right)^{2}}+\frac{1}{2}, & q_0w\longrightarrow\infty.
\end{array}\right.
\ee
As an example, the total variational energy $U_{\text{var}}+K_{\text{var}}$, is plotted as a function of $\eta$ in Fig. \ref{fig: TE}.
